\documentclass[aps,prl,twocolumn,groupedaddress]{revtex4}

\usepackage{amsmath}
\usepackage{amssymb}
\usepackage{amsfonts}
\usepackage{graphicx}

\newcommand{\Ref}{Ref.\,}
\newcommand{\Refs}{Refs.\,}
\newcommand{\Fig}{Fig.\,}
\newcommand{\Figs}{Figs.\,}
\newcommand{\Tab}{Tab.\,}

\newcommand{\Eq}{Eq.\,}
\newcommand{\Eqs}{Eqs.\,}
\newcommand{\B}{\mathrm{Born}}
\newcommand{\Oc}{\mathcal{O}}
\newcommand{\ds}{\displaystyle}
\newcommand{\cq}{\textit{quasiclassical }}
\newcommand{\Du}{Delbr\"uck }
\newcommand{\BA}{lowest-order Born approximation}
\newcommand{\CC}{Coulomb corrections }
\newcommand{\cci}{\textit{conditionally convergent iterated} }
\begin{document}


\title{Coulomb corrections to the \Du scattering amplitude at low
energies}


\author{G.G. Kirilin}
\email[]{G.G.Kirilin@inp.nsk.su}
\affiliation{Budker Institute of Nuclear Physics, SB RAS,
Novosibirsk}

\author{I.S. Terekhov}
\email[]{I.S.Terekhov@inp.nsk.su}
\affiliation{Budker Institute of Nuclear Physics, SB RAS,
Novosibirsk}

\date{\today}

\begin{abstract}
In this article, we study the \CC to the \Du scattering amplitude.
We consider the limit when the energy of the photon is much less
than the electron mass. The calculations are carried out in the
coordinate representation using the exact relativistic Green
function of an electron in a Coulomb field. The resulting relative
corrections are of the order of a few percent for scattering on
for a large charge of the nucleus. We compare the corrections with
the corresponding ones calculated through the dispersion integral
of the pair production cross section and also with the magnetic
loop contribution to the g-factor of a bound electron. The last
one is in a good agreement with our results but the corrections
calculated through the dispersion relation are not.
\end{abstract}


\maketitle

\section{Introduction \label{s1}}



The elastic scattering of photons by an external Coulomb field
(the so called \Du scattering \cite{Meitner1933}) is one of
nontrivial predictions of quantum electrodynamics. In the
perturbation theory, the \Du scattering amplitude begins from the
second order in $Z \alpha$ ($Z|e|$ is the charge of the nucleus,
$\alpha = e^2 \approx 1/137$ is the fine-structure constant, we
put $c=1$, $\hbar=1$). Therefore, significant efforts have been
made to calculate the amplitude for the arbitrary scattering
angles and energies even in the \BA. The results of these
calculations and the detailed bibliography can be found in the
report \cite{Papatzacos:1975bu}.

To calculate the \Du scattering amplitude for $Z\gg 1$ it is
necessary to take into account Coulomb field exactly. The
expression for the amplitude exact in $Z\alpha$ has been derived
in \Ref\cite{Scherdin:1991hr} without any additional assumptions,
but numerical results have not yet been obtained because it is
fairly cumbersome.

Considerable progress in the calculation of the \CC to the \BA\
has been achieved for the case of the photon energy $\omega$ much
large than the electron mass $m_e$ and either at small scattering
angles $\Delta/\omega\ll~1$ ($\Delta=|\mathbf{k}_1-\mathbf{k}_2|$,
where $\mathbf{k}_1$ and $\mathbf{k}_2$ are the momenta of the
photon in the initial and final states, correspondingly)
\Refs\cite{Cheng:1969ab, Cheng:1971vw, Cheng:1972zu, Milstein1983,
Milstein1983a} or at large momentum transfer $\Delta/m_e\gg~1$
\Refs\cite{Cheng:1982xs, Milstein1988}. It turns out, that the \CC
strongly decrease the \Du amplitude in comparison with the \BA\
(the theoretical results and the corresponding experimental data
are reviewed in detail in the report \cite{Milstein1994} and the
tables \cite{Falkenberg1992}). At the moment, the minimal photon
energy at which \Du scattering is experimentally observed is
$\omega=889\,\mathrm{KeV}$
\Refs\cite{Schumacher1969,Smend1973,Smend1973a,Smend1974,Muckenheim1980}
but the corresponding energy for the \CC is
$\omega=2754\,\mathrm{KeV}$ \Ref\cite{Rullhusen1979} only, i.\,e.
above the electron-positron pair production threshold.

We aim here to calculate the \CC at low energy $\omega \ll m_e$.
These corrections have not yet been investigated neither
experimentally nor theoretically. Nevertheless, they are closely
connected with the \CC to the pair production cross section due to
the dispersion relation \Refs\cite{Rohrlich1952, Rohrlich1957}
and, as we will show below, with the magnetic-loop contribution to
the $g$-factor of a bound electron \Ref\cite{Beier2000}. We think
that the \Du scattering amplitude at low energy is useful to
estimate the \CC for both phenomena.

The amplitude is calculated in the coordinate representation with
the help of the Green function of an electron moving in a Coulomb
field. The structure of this paper is the following: in Section 2,
we provide all necessary information about the Green function and
the general parametrization of the \Du scattering amplitude. In
Section 3, we show that the calculation in the coordinate
representation reproduces the result of the \BA\ derived in the
momentum representation. We point out the difficulties specific
for a calculation in the coordinate representation also occurring
during the calculation of the \CC\!\!. The  results for the \CC
are given in Section 4. We also provide the simple parametrization
of their dependence on $Z$. In Section 5 we compare our results
with those obtained via the dispersion relation. The estimated
value for the magnetic-loop contribution to the $g$-factor of a
bound electron is given in Section 6.

\section{\Du scattering amplitude \label{s2}}

We parameterize the \Du scattering amplitude as follows:
\begin{align}
A =\epsilon_{\mu}^{(1)}\,\epsilon_{\nu}^{\star(2)}\,\Pi^{\mu\nu}\left(  \omega,\mathbf{k}_{1},\mathbf{k}%
_{2},Z\right),\label{e2:1}
\end{align}
\begin{multline}
\Pi^{\mu\nu}\left(  \omega,\mathbf{k}_{1},\mathbf{k}_{2},Z\right)= \\
= \frac{\alpha\left(  Z\alpha\right)  ^{2}}{m_{e}^{3}}\left\{
f_{1}\left( \omega,\mathbf{k}_1,\mathbf{k}_2,Z\right)  \,\left[
g^{\mu\nu}\,k_{1}\cdot k_{2}-k_{2}^{\mu
}k_{1}^{\nu}\right]  \right. \\
\left.  +f_{2}\left(  \omega,\mathbf{k}_1,\mathbf{k}_2,Z\right)
\,\left[ \omega^{2}g^{\mu\nu
}-\omega\left(  n^{\mu}k_{1}^{\nu}+n^{\nu}k_{2}^{\mu}\right)\right.\right.\\
\left. \left. +\,n^{\mu}%
n^{\nu}\,k_{1}\cdot k_{2}\right]  \right\}\label{e2:2}
\end{multline}
where $k_1=(\omega,\mathbf{k}_1), k_2=(\omega,\mathbf{k}_2)$ are
the 4-momenta of the photon in the initial and final states,
correspondingly, $\epsilon^{(1,2)}$ are the polarization vectors,
the 4-vector $n$ is defined as $k_1\cdot n=k_2\cdot n = \omega$,
$f_{1}$ and $f_{2}$ are the form factors to be calculated. The
main purpose of this article is to calculate
$\lim_{|\mathbf{k}|\to
0}f_{1,2}\left(0,\mathbf{k},\mathbf{k},Z\right)$.

In a point-like charge approximation (Coulomb field), the
polarization tensor $\Pi^{\mu\nu}$ has the following form:
\begin{align}
& \Pi^{\mu\nu}\left(\ldots,Z\right)
=\tilde {\Pi}^{\mu\nu}\left(
\ldots,Z\right)
-\tilde{\Pi}^{\mu\nu}\left(\ldots,0\right),\label{e2:3}\\
& \tilde{\Pi}^{\mu\nu}( \omega,\mathbf{k}_{1},\mathbf{k}_{2},Z)
=i\alpha\int d^{3}r_{1}d^{3}r_{2}\exp(i\mathbf{k}_{1}\mathbf{r}%
_{1}-i\mathbf{k}_{2}\mathbf{r}_{2})  \notag\\
& \times\int_{C}\frac{d\epsilon}{2\pi}\,\mathrm{Sp}\left\{
\gamma^{\mu}\hat{G}\left(
\mathbf{r}_{1},\mathbf{r}_{2}|\epsilon\right)  \gamma^{\nu
}G\left(  \mathbf{r}_{2},\mathbf{r}_{1}|\epsilon-\omega\right)
\right\},\label{e2:4}
\end{align}
where $\hat{G}\left(\mathbf{r}_{1},\mathbf{r}_{2}|\epsilon\right)$
is the Green function of an electron in a Coulomb field. The
contour of integration over $\epsilon$ in the expression
(\ref{e2:4}) goes from $-\infty$ to $\infty$ so that it is below
the real axis in the left half-plane and above the real axis in
the right half-plane. It is convenient to turn the contour along
the imaginary axis. In this case the Green function takes the
following form (see \Ref\cite{Milshtein1982}):
\begin{widetext}
\begin{align}
G(\mathbf{r}_{1},\mathbf{r}_{2}  & \mid i\epsilon)
=\frac{-1}{4\pi r_{1}r_{2}\,p}\sum_{l=1}^{\infty}\int_{0}^{\infty}%
ds \exp(2iZ\alpha\,s\,(\epsilon/p)-p(r_{1}+r_{2})\coth s)\notag\\
& \times\left\{\left[  R_{+}\frac{y}{2}\,I_{2\nu}^{\prime}%
(y)B_{l}+R_{-}lI_{2\nu}(y)A_{l}\right]  \left(
\gamma_{0}\,i\epsilon
+m\right)
+Z\alpha\gamma^{0}\left[  im\left(  \mathbf{\hat{n}}_{1}+\mathbf{\hat{n}%
}_{2}\right)+ p\,R_{+}\,\coth s\right]  I_{2\nu}\left(  y\right)  B_{l}\right.\notag\\
& \left.-i\left[  p^{2}\frac{r_{1}-r_{2}}{2\sinh^{2}s}\,\left(  \mathbf{\hat{n}}%
_{1}+\mathbf{\hat{n}}_{2}\right)  B_{l}+p\coth s\left(  \mathbf{\hat{n}}%
_{1}-\mathbf{\hat{n}}_{2}\right)  lA_{l}\right]
I_{2\nu}(y)\right\}, \label{e2:9}
\end{align}
where
\begin{align}
R_{\pm} & =1\pm\mathbf{n}_{1}\,\mathbf{n}_{2}\pm
i\mathbf{\Sigma}\,\left(
\mathbf{n}_{1}\times\mathbf{n}_{2}\right),\qquad \Sigma^{k}
=i\epsilon^{ijk}\left[  \gamma^{i},\gamma^{j}\right]/4,\qquad
\mathbf{\hat{n}}_{(1,2)}=\boldsymbol{\gamma\,n}_{(1,2)},\notag\\
A_{l} & =\frac{d}{dx}\left( P_{l}(x)+P_{l-1}(x)  \right),\qquad
B_{l} =\frac{d}{dx}\left(P_{l}(x)-P_{l-1}(x)\right),\qquad
x =\mathbf{n}_{1}\,\mathbf{n}_{2},\notag\\
\nu & =\sqrt{l^{2}-\left(  Z\alpha\right)^{2}},\qquad y =2p\sqrt{r_{1}r_{2}}/\sinh s,\qquad p =\sqrt{m^{2}+\epsilon^{2}%
}. \label{e2:10}
\end{align}
\end{widetext}
For the sake of convenience, we calculate the time-time component
of the polarization tensor and the trace of the spatial components
separately:
\begin{align}
\Pi^{(ii)}\left(\mathbf{k},Z\right)   =\left(  n_{\mu}%
n_{\nu}-g_{\mu\nu}\right)  \Pi^{\mu\nu}\left(
0,\mathbf{k},\mathbf{k},Z\right),\label{e2:5}
\end{align}
\begin{align}
\Pi^{(00)}\left(\mathbf{k},Z\right)   =n_{\mu}n_{\nu}%
\Pi^{\mu\nu}\left(\mathbf{k},Z\right)\label{e2:6}.
\end{align}
Substituting the parametrization of $\Pi^{\mu\nu}$ (\ref{e2:2}) in
the right-hand side of the \Eqs(\ref{e2:5}) and (\ref{e2:6})
yields the relation between $\Pi^{(ii)}, \Pi^{(00)}$ and the form
factors $f_{(1,2)}$:
\begin{align}
\Pi^{(ii)}\left(\mathbf{k},Z\right) =2\frac{\alpha\left(
Z\alpha\right)^{2}}{m_{e}^{3}}\,\mathbf{k}^{2}f_{1}\left(  0,\mathbf{k}%
,\mathbf{k},Z\right)\label{e2:7}
\end{align}
\begin{align}
\Pi^{(00)}\left(\mathbf{k},Z\right)& =-\frac{\alpha (Z\alpha)
^{2}}{m_{e}^{3}}\,\mathbf{k}^{2}\left(f_{1}(0,\mathbf{k},\mathbf{k},Z)\right.\notag\\&
\left.+f_{2}(0,\mathbf{k},\mathbf{k},Z)\right).\label{e2:8}
\end{align}

\section{\BA \label{s3}}
\begin{figure}
  \includegraphics[width=.35\textwidth]{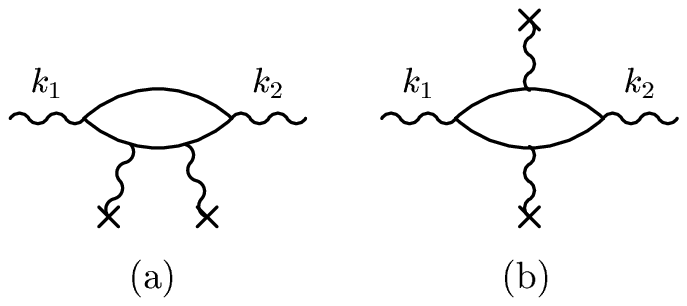}\\
  \caption{\BA}\label{f3:1}
\end{figure}
The diagrams of the second order of the perturbation theory in
$Z\alpha$ are depicted in \Fig\ref{f3:1}. Their contribution were
calculated in \Refs\cite{Costantini:1971cj,Papatzacos:1975bt}. We
aim here to demonstrate that the calculation of this diagrams in
the coordinate representation reproduces the result of the \BA.

To calculate the contributions of the diagrams, see
 \Fig\ref{f3:1}, we expand the exponent function in the expression
(\ref{e2:4}) up to the second order in
$|\mathbf{k}_1|=|\mathbf{k}_2|=|\mathbf{k}|$:
\begin{align}
&2 \Pi^{\mu\nu}_{(a)} + \Pi^{\mu\nu}_{(b)}
=\frac{\alpha\,\mathbf{k}^2}{6}\int d^{3}r_{1}d^{3}r_{2}%
|\mathbf{r}_{1}-\mathbf{r}_{2}|^2\int_{-\infty}^{\infty}\frac{d\epsilon}{2\pi}  \notag\\
& \times \mathrm{Sp}\left\{ 2\,\gamma^{\mu}G^{(0)}\left(
\mathbf{r}_{1},\mathbf{r}_{2}|i\epsilon\right)  \gamma^{\nu}
G^{(2)}\left( \mathbf{r}_{2},\mathbf{r}_{1}|i\epsilon\right) \right.\notag\\
&\left. + \gamma^{\mu}G^{(1)}\left(
\mathbf{r}_{1},\mathbf{r}_{2}|i\epsilon\right)  \gamma^{\nu}
G^{(1)}\left( \mathbf{r}_{2},\mathbf{r}_{1}|i\epsilon\right)
\right\},\label{e3:4}
\end{align}
where $G^{(n)}$ is the contribution to the Green function
(\ref{e2:9}) of the $n$-th order in~$Z\alpha$:
\begin{align}
G^{(0)}& =-\left[ \left(
\gamma_{0}\,i\epsilon+m\right)  +i\,\frac{ \mathbf{\hat{r}}_{1}%
-\mathbf{\hat{r}}_{2}}{\left\vert
\mathbf{r}_{1}-\mathbf{r}_{2}\right\vert}
\frac{\partial}{\partial\left\vert
\mathbf{r}_{1}-\mathbf{r}_{2}\right\vert}\right]\notag\\
&\times\frac{\exp\left( -\left\vert
\mathbf{r}_{1}-\mathbf{r}_{2}\right\vert \right)  }{4\pi\left\vert
\mathbf{r}_{1}-\mathbf{r}_{2}\right\vert },\label{e3:5}
\end{align}
\begin{widetext}
\begin{align}
G^{(1)}& =-\frac{Z\alpha}{16\pi}\frac{1-t^{2}}{r_{1}r_{2}}\left\{
2i\frac{\epsilon\rho}{p^{2}}\left[ \gamma_{0}\,i\epsilon+ m
+\frac{ip^{2}}{\rho^{2}}\left(
\mathbf{\hat{r}}_{1}-\,\mathbf{\hat{r}}_{2}\right)
\dfrac{\partial}{\sigma\partial\sigma }+\frac{ip}{2}\left(
\mathbf{\hat{n}}_{1}-\mathbf{\hat{n}}_{2}\right)
\dfrac{\partial}{\partial\rho}\right]F(\rho,\sigma)\right. \notag \\
& +\left.  \dfrac{1}{1-\sigma^{2}}\,\gamma^{0}\left(  \hat{R}_{+}-i\frac{m}%
{p}\,\left(  \mathbf{\hat{n}}_{1}+\mathbf{\hat{n}}_{2}\right)
\frac{\partial }{\partial\rho}\right)  \left(
\dfrac{1}{\,\sigma}e^{-\rho\,\sigma}-e^{-\rho}\right)  \right\}  ,\label{e3:6}\\
G^{(2)}& =\frac{(Z\alpha)^{2}}{4\pi
r_{1}r_{2}}\int_{0}^{\infty}ds\,\exp(ip(r_{1}+r_{2})\coth s)\left\{
\sum_{l=1}^{\infty}\left[\left(\hat{R}_{+} B_{l}\,\frac{y\,\partial}{2\,\partial y}
+\hat{R}_{-} l A_{l}\right)(\gamma_{0}\,i\epsilon/p + m/p) \right.\right.\label{e3:7}\\
& -\left.i\,\frac{p\,(r_{1}-r_{2})}{2\sinh^{2}s}\left(
\mathbf{\hat{n}}_{1}+\mathbf{\hat{n}}_{2}\right) B_{l}- i \coth s\left(
\mathbf{\hat{n}}_{1}-\mathbf{\hat{n}}_{2}\right) lA_{l}\right]\left.\frac{\partial}{4l\,\partial\nu}
I_{2\nu}(y)\right\vert_{\nu=l}\label{e3:8}\\
& \left. + \frac{\epsilon s}{2p}\left[  \frac{\epsilon s}{2p}\,\hat{T}+\gamma^{0}\left(
\frac{m}{p}\left(  \mathbf{\hat{n}}_{1}+\mathbf{\hat{n}}_{2}\right)  -\hat
{R}_{+} i\coth s\right)\,\frac{y}{\cos\frac{\phi}{2}}\,I_{1}\left(
y\cos\frac{\phi}{2}\right)  \right] \right\},\label{e3:9}
\end{align}
where
\begin{align}
F(\rho,\sigma)  & =\frac{1}{2\rho\sigma}\left[
e^{-\rho\,\sigma}\log
\dfrac{1+\sigma}{1-\sigma}+e^{\rho\,\sigma}\,\Gamma\left(  0,\rho
\,(1+\sigma)\right)  -e^{-\rho\,\sigma}\,\Gamma\left(
0,\rho\,(1-\sigma
)\right)  \right] ,\\
\hat{T}  & =\left[ 2\,y^{2}\,\left(  \gamma_{0}\,i\epsilon/p +
m/p\right) -i\,y^{2}\coth s\,\left(
\mathbf{\hat{n}}_{1}-\mathbf{\hat{n}}_{2}\right)
\,-i\,\frac{p\,(r_{1}-r_{2})}{\sinh^{2}s}\,\frac{\left(  \mathbf{\hat{n}}%
_{1}+\mathbf{\hat{n}}_{2}\right)
}{\cos^{2}\frac{\phi}{2}}\,\frac{y\,\partial }{\partial y}\right]
I_{0}\left(  y\cos\frac{\phi}{2}\right) \label{e3:10} ,
\end{align}
\end{widetext}
where $\ \Gamma\left(  a,z\right)  =\int_{z}^{\infty}
t^{a-1}\exp\left(  -t\right)  \,dt$ is the incomplete gamma
function. The following notations have been also introduced in the
expressions (\ref{e3:5}-\ref{e3:10}):
\begin{align}
\rho&=p\left(  r_{1}+r_{2}\right), & \sigma&=\frac{\left\vert \mathbf{r}%
_{1}-\mathbf{r}_{2}\right\vert }{r_{1}+r_{2}},
\\ t&=\frac{r_{1}-r_{2}}
{r_{1}+r_{2}},& \cos\frac{\phi}{2}&=\left(  \frac{1+\mathbf{n}_{1}%
\mathbf{n}_{2}}{2}\right)  ^{1/2}.\label{e3:11}
\end{align}
The contribution to the Green function $G^{(2)}$ of the second
order is split into two parts: the contribution corresponding to
the expansion of a Bessel function over an index (the part of the
expressions (\ref{e3:7}), (\ref{e3:8}) separated by the square
brackets) and the contribution in which we substitute the indices
of the Bessel functions for $2l$ and sum over $l$ explicitly. The
latter can be called conditionally \textit{quasiclassical}
contribution because it corresponds to the contribution of the
large angular momenta $l\gg Z\alpha$.

\begin{figure}[t]
\includegraphics[width=0.47\textwidth]{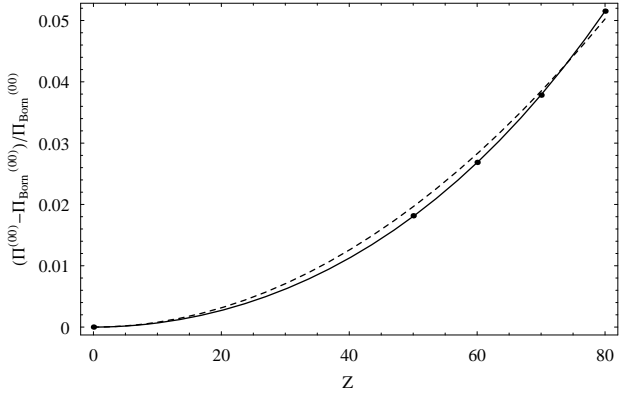}
\caption{ The relative \CC to the trace of the time-time component
of the polarization tensor. The dashed curve corresponds to the
fit $a (Z\alpha)^2$, the solid line corresponds to $a
(Z\alpha)^2+b (Z\alpha)^4$}\label{f4:1}
\end{figure}

There are some points to be made. First of all, the contribution
of each diagram \Fig\ref{f3:1} is infrared divergent, i.\,e. it is
divergent at large distances. Since the divergence is canceled
between the diagrams \Fig\ref{f3:1}(a) and \Fig\ref{f3:1}(b), the
contribution each of them depends on the regularization of this
divergence. However, the contribution of separated terms (for
example, the \textit{quasiclassical} contribution or the
contribution proportional to $\partial_\nu I_{2\nu}$ (\ref{e3:7}),
(\ref{e3:8})) and even the presence of the divergence depends on
the order of the iterated integration over the spatial variables
$\mathbf{r}_i$ and the inner variables of the Green functions --
"proper times" $s_i$. An iterated improper integral, the value of
which depends on the order of integration, we call a \cci
integral. The example of such integral, that arises during the
calculation of the diagram \Fig\ref{f3:1}(b), is given in
Appendix. To avoid the complications due to an explicit
regularization and difficulties during the calculation of
separated terms, one should fix the order of integration for all
the diagrams and sum the contribution of each one before the
integration with respect to the last variable. We chose the
variable $t$ defined in (\ref{e3:11}) as the last integration
variable. In this case, the contributions of the diagrams
(\Fig\ref{f3:1}) have the following form:
\begin{align}
\Pi^{(ii)}_{(a)} = \frac{\alpha (Z\alpha)
^{2}}{m_{e}^{3}}\,\mathbf{k}^{2} \left( -\frac{5}{2304} -
\frac{5}{128}\int^1_0\frac{dt}{t^2}\right),
\end{align}
\begin{align}
\Pi^{(ii)}_{(b)} = \frac{\alpha
(Z\alpha)^{2}}{m_{e}^{3}}\,\mathbf{k}^{2} \left( \frac{19}{1152} +
\frac{5}{64}\int^1_0\frac{dt}{t^2}\right),\label{e3:12}
\end{align}
\begin{align}
\Pi^{(ii)}_\B = 2\Pi^{(ii)}_{(a)} + \Pi^{(ii)}_{(b)}  =
\frac{\alpha\left( Z\alpha\right)^{2}}{m_{e}^{3}}\,\mathbf{k}^{2}
\frac{7}{576}. \label{e3:1}
\end{align}
The time-time component is derived in a similar manner:
\begin{align}
\Pi^{(00)}_\B  = \frac{\alpha (Z\alpha)
^{2}}{m_{e}^{3}}\,\mathbf{k}^{2} \frac{59}{2304}.\label{e3:2}
\end{align}
Substituting the expressions (\ref{e3:1}) and (\ref{e3:2}) in
(\ref{e2:7}) and (\ref{e2:8}) yields the form factors (\ref{e2:4})
in the \BA
\begin{align}
f_{1\,\mathrm{B}}=\frac{7}{16\cdot
72},\:f_{2\,\mathrm{B}}=-\frac{73}{32\cdot 72}\,.\label{e3:3}
\end{align}
As noticed above, they coincide with the results derived in
\cite{Costantini:1971cj,Papatzacos:1975bt}.

\section{\CC \label{s4}}

\begin{figure}[t]
\includegraphics[width=0.47\textwidth]{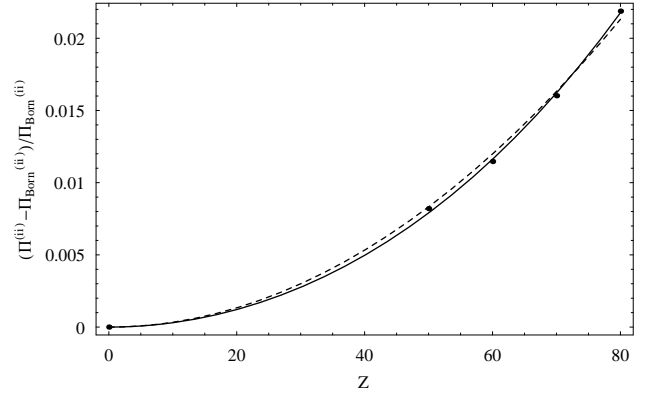}
\caption{The relative \CC to the trace of the spatial components
of the polarization tensor. The dashed curve corresponds to the
fit $a (Z\alpha)^2$, the solid line corresponds to $a
(Z\alpha)^2+b (Z\alpha)^4$}\label{f4:2}
\end{figure}

Analytical derivation of the \CC to the \BA\ (\ref{e3:3}) is a
rather complicated problem.
We have calculated these corrections mostly numerically. To
increase the accuracy of the numerical calculations we have
subtracted the \BA\  \Eq(\ref{e3:4}) from the general expression
\Eq(\ref{e2:4}) before any transformations.
Let us denote the angular momenta and the proper times that appear
in the Green functions in \Eq(\ref{e2:4}) as $l_1$, $l_2$ and
$s_1$, $s_2$. After subtraction, it is convenient to change
variables as follows: $(r_1,r_2)\to(\eta=\sqrt{r_1 r_2}, t' =
r_1/r_2)$. Then we integrate analytically over $\epsilon$,
$x=(\mathbf{r}_1\mathbf{r}_2)/|\mathbf{r}_2||\mathbf{r}_1|$, $t'$
one by one. After that we also perform the analytical summation
over $l_1$.
Thus, the expression (\ref{e2:4}) can be reduced to the sum over
$l_2$ and the iterated integral over $s_1$, $s_2$ and $\eta$. The
explicit expression for the integrand is omitted here as bulky.
Further analytical integration is only possible for separate
terms.
The term previously referred to as the \textit{quasiclassical}
(which contains $I_{2l}$ instead of $I_{2\nu}$) can be represented
as an one-fold integral or as an infinite series over $Z\alpha$.
For example, the corresponding contribution to $\Pi^{(00)}$ is the
following :
\begin{align}
&  \Pi_{\mathrm{quasicl.}}^{(00)}-\Pi_{\mathrm{quasicl.Born}}^{(00)}%
=\frac{\alpha\left(  Z\alpha\right)
^{2}}{m_{e}^{4}}\,\mathbf{k}^{2}\notag\\
&\times\left( \frac{287\pi^{2}-39}{18432}\left(  Z\alpha\right)
^{2}\right. -\frac{\pi^{2}\left(  49\pi^{2}-15\right)
}{69120}\left( Z\alpha \right)^{4}\notag\\
&+\frac{\pi^{4}\left( 158\pi^{2}-63\right) }{3096576}\left(
Z\alpha\right)^{6}  -\frac{\pi^{6}\left( 21\pi^{2}-10\right)
}{5160960}\left( Z\alpha\right)
^{8}\notag\\
&+\left.\frac{\pi^{8}\left( 83290\pi^{2}-46431\right)
}{245248819200}\left(  Z\alpha\right)
^{10}+\ldots\right)\label{e4:1}.
\end{align}
This contribution is finite, i.\,e. the infrared divergency is
absent in \Eq(\ref{e4:1}). Nevertheless, the residual part must be
integrated in the same order as that used to derive
\Eq(\ref{e4:1}). We check explicitly that the contribution
containing the subtraction from the single Bessel function, i.\,e.
which is proportional to
\begin{align}
I_{2\nu}-I_{2l}-\left.\frac{(Z\alpha)^2}{2l}\,\partial_{\nu}I_{2\nu}\right|_{\nu=l}
= O(Z^4\alpha^4),
\end{align}
is a \cci integral.

The results of our numerical calculations are presented in
\Figs\ref{f4:1},\:\ref{f4:2} and \Tab\ref{t4:1}.

\begin{table}
\caption{Relative \CC \label{t4:1}}
\begin{ruledtabular}
\begin{tabular}
[c]{ccc} Z &
$\left(\Pi^{(00)}-\Pi_{\mathrm{Born}}^{(00)}\right)/\Pi_{\mathrm{Born}}^{(00)}$
& $\left(\Pi^{(ii)}-\Pi_{\mathrm{Born}}^{(ii)}\right)/\Pi_{\mathrm{Born}}^{(ii)}%
$\\\hline
$50$ & $ 1.82 \cdot10^{-2}$ & $8.20\cdot10^{-3}$\\
$60$ & $ 2.69\cdot10^{-2}$ & $1.15\cdot10^{-2}$\\
$70$ & $ 3.78\cdot10^{-2}$ & $1.60\cdot10^{-2}$\\
$80$ & $ 5.15\cdot10^{-2}$ & $2.19\cdot10^{-2}$\\
\end{tabular}
\end{ruledtabular}
\end{table}

It should be noted that the \cq contribution is of the order of
$60\%$ of the \BA\ for $Z=80$. The residual part containing the
substraction from Bessel functions reduces the corrections up to a
few percents. This cancellation adversely affects the accuracy of
the calculation. The authors carried out the calculations of the
corrections independently. The accuracy of the results
(\Tab\ref{t4:1}) is better than one percent.


%
Now we consider the dependence of the Coulomb corrections on $Z$.
The first two terms in \Eq(\ref{e4:1}) dominate and give $97-96\%$
of the \cq contribution in spite of the fact that we calculate
complete series in $Z\alpha$. It comes as a surprise that the
complete results can be adequately fitted by a biquadratic
polynomial in $Z\alpha$ without a free term:
\begin{align}
\frac{\Pi^{(00)}-\Pi_{\mathrm{Born}}^{(00)}}{\Pi_{\mathrm{Born}}^{(00)}}
& =\frac{2304}{59%
}\left[ \left(  3.22\pm0.01\right) \cdot10^{-3}\left(  Z\alpha
\right)  ^{2} \right. \notag\\
&\left.+\left(  1.90\pm0.02\right)  \cdot10^{-3}\left(
Z\alpha\right) ^{4}\right]  ,\label{e4:2}
\end{align}
\begin{align}
\frac{\Pi^{(ii)}-\Pi_{\mathrm{Born}}^{(ii)}}{\Pi_{\mathrm{Born}}^{(ii)}}& =\frac{576}{7%
}\left[\left(  6.69\pm0.17\right) \cdot10^{-4}\left(  Z\alpha
\right)^{2}\right.\notag\\
& \left.+\left(  3.18\pm0.54\right)\cdot10^{-4}\left(
Z\alpha\right) ^{4}\right]  \label{e4:3}.
\end{align}
The results of the fitting with a quadratic function $a\,
(Z\alpha)^2$ and also the functions in
\Eqs(\ref{e4:2}),\:(\ref{e4:3}) are shown in
\Figs\ref{f4:1},\:\ref{f4:2}.
One further comment is in order. The coefficients at $(Z\alpha)^2$
in \Eqs(\ref{e4:2}) and (\ref{e4:3}) have a magnitude one or two
orders less than those at $(Z\alpha)^0$ in the \BA\
\Eqs(\ref{e3:1}),\:(\ref{e3:2}). If one assumes the same hierarchy
between the coefficients at $(Z\alpha)^2$ and $(Z\alpha)^4$, then
the coefficients of $(Z\alpha)^4$ could not be distinguished from
zero with our accuracy.
In this case, the maximal difference between the dashed and solid
curves in \Figs\ref{f4:1},\:\ref{f4:2} shows the actual accuracy
of our calculations.
Substituting \Eqs(\ref{e4:2}), (\ref{e4:3}) in the relations
\Eqs(\ref{e2:7}), (\ref{e2:8}), we obtain the form factors
$f_{(1,2)}$:
\begin{align}
f_{1} =\frac{7}{1152}+3.35\cdot10^{-4}(Z\alpha)^{2}%
+1.6\cdot10^{-4}(Z\alpha)^{4}, \label{e4:4}%
\end{align}
\begin{multline}
f_{2} =-\frac{73}{2304}-3.36\cdot10^{-3}(Z\alpha)^{2}%
\\-2.1\cdot10^{-3}(Z\alpha)^{4}. \label{e4:5}%
\end{multline}

\section{Pair production via dispersion relation \label{s5}}

\begin{figure}
\includegraphics[width=0.47\textwidth]{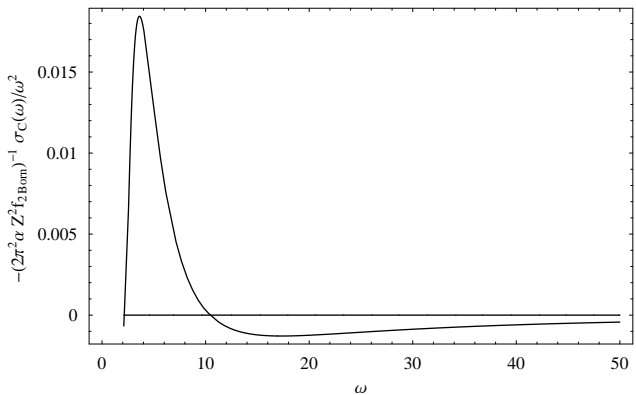}
\caption{\CC to the pair production cross section $(Z=82)$}
\label{f5:1}
\end{figure}

In the papers \cite{Rohrlich1957}, \cite{Rohrlich1952}, Gluckstern
and Rohrlich have derived the relation between the pair production
cross section in a Coulomb field and the \Du  amplitude averaged
over the polarizations:
\begin{align}
A(\omega) & =\frac{\omega^{2}}{2\pi^{2}}\int
_{2m}^{\infty}\frac{\sigma_{\gamma\rightarrow e^{+}e^{-}}\left(
\omega^{\prime}\right)
}{\omega^{\prime2}-\omega^{2}+i0}\,d\omega^{\prime}. \label{e5:5}
\end{align}
The amplitude (\ref{e2:1}) averaged over the polarizations of the
photon has the form:
\begin{equation}
A=\frac{1}{2}\left(
\delta^{ij}-\frac{k^{i}k^{j}}{\omega^{2}}\right)
\Pi^{ij}=-\frac{\alpha\left(  Z\alpha\right)  ^{2}}{m_{e}^{3}}f_{2}%
\,\omega^{2}. \label{e5:6}%
\end{equation}
One can find the relation between the \CC to the form factor
$f_{2}$ and the pair production cross section in a Coulomb field
by using the expressions (\ref{e5:5}), (\ref{e5:6}) (we put
$m_e=1$ in this section):
\begin{equation}
f_{2}\,=-\frac{1}{2\pi^{2}\alpha\left(  Z\alpha\right)
^{2}}\int_{2}^{\infty
}\frac{\sigma\left(  \omega^{\prime}\right)  \,}{\omega^{\prime2}}%
\,d\omega^{\prime}. \label{e5:7}%
\end{equation}
Let us check the formula (\ref{e5:7}) in the Born approximation.
Substituting $Z=82$ (lead) yields:
\begin{equation}
\frac{1}{2\pi^{2}\alpha\left(  Z\alpha\right)
^{2}f_{2B}}\int_{2}^{\infty
}\frac{-\sigma_{B}\left(  \omega^{\prime}\right)  \,}{\omega^{\prime2}%
}\,d\omega^{\prime}=1+4\cdot10^{-5},
\end{equation}
where $\sigma_{B}$ is replaced by the asymptotical formulae
derived by Maximon in \Ref\cite{Maximon1968} for $\omega<2.1$:
\begin{align}
&\sigma_{B}(\omega)=\alpha\left(  Z\alpha\right) ^{2}
\frac{2\pi}{3}\left(  \frac{\omega-2}{\omega}\right)
^{3}\notag\\
&\times\left(
1+\frac{\epsilon}{2}+\frac{23}{40}\,\epsilon^{2}+\frac{11}{60}\,\epsilon
^{3}+\frac{29}{960}\epsilon^{4}+\Oc(\epsilon^{5})\right),\\
& \epsilon =\frac{2\omega-4}{2+\omega+2\,(2\omega)^{1/2}},
\end{align}
for $\omega>2.1$:
\begin{align}
& \sigma_{B}\left(  \omega\right)  =  \alpha\left(
Z\alpha\right)^{2}\left\{\frac{28}{9}\log2\omega-\frac{218}{27}\right.\notag\\
&\left.+\left( \frac{2}{\omega}\right) ^{2}\left[
6\log2\omega-\frac{7}{2}+\frac{2}{3}\log^{3}2\omega-\log
^{2}2\omega\right.\right.\notag\\
&\left.\left.-\frac{\pi^{2}}{3}\log2\omega+\frac{\pi^{2}}{6}+2\zeta\left(
3\right)  \right]  -\left(  \frac{2}{\omega}\right)  ^{4}\left[  \frac{3}{16}%
\log2\omega+\frac{1}{8}\right]  \nonumber\right.\\
&\left.   -\left(  \frac{2}{\omega}\right) ^{6}\left[
\frac{29\log2\omega}{9\cdot256}-\frac{77}{27\cdot512}\right] +\Oc
\left( \frac{2^8}{\omega^8}\right)\right\} .
\end{align}
Now we discuss the Coulomb corrections to the form factors. Using
the \Eq(\ref{e4:5}) we have obtained the relative correction to
the form factor in the Born approximation (here and below all the
calculations are carried out for $Z=82$):
\begin{equation}
\frac{f_{2}-f_{2B}}{f_{2B}}=4.9\cdot10^{-2}. \label{e5:8}%
\end{equation}
However, if we use the \CC to the pair production cross section
$\sigma_{C}\left(  \omega\right)  =\sigma\left( \omega\right)
-\sigma _{B}\left(  \omega\right)  $ derived in
\Ref\cite{Overbo1973} for the photon energy $\omega<10$ and the
interpolation equation derived in \Ref\cite{Overbo:1977mh} for
$\omega>10$, then  the relative correction to the form factor in
the Born approximation is
\begin{equation}
-  \frac{1}{2\pi^{2}\alpha\left(  Z\alpha\right) ^{2}f_{2B}}\int
_{2.01}^{\infty}\frac{\sigma_{C}\left(  \omega\right)  \,}{\omega^{2}}%
\,d\omega =2.7\cdot10^{-3}. \label{e5:9}%
\end{equation}
This result is 20 times less than that in \Eq(\ref{e5:8}). The
integrand (\ref{e5:9}) as a function of $\omega$ is shown in
\Fig\ref{f5:1}.
It varies mainly in the region $2<\omega<30$ but there is a long
negative "tail" for $\omega\to\infty$. The total integral is a
result of the almost complete cancelation between the positive
contribution for $\omega\lesssim10$ and the negative one for
$\omega\gtrsim10$. The following ratio shows it clearly:
\begin{equation}
\frac{\int_{2}^{\infty}\sigma_{T }\left(  \omega\right)  /\omega
^{2}\,d\omega}{\int_{2}^{\infty}\left\vert \sigma_{T }\left(
\omega\right)  \,/\omega^{2}\right\vert
\,d\omega}=3.9\cdot10^{-2}.\label{e5:12}
\end{equation}
For the integral (\ref{e5:9}) to be calculated with the sufficient
accuracy it is necessary to derive the \CC to the cross section
with an accuracy better than a few percents. It is quite possible
that this cancelation causes the discrepancy due to the lack of
precision in the calculations of the positive part of the
integrand in Ref.\cite{Overbo1973}.

The cause of the discrepancy could also be the interpolation
equation derived in \Ref\cite{Overbo:1977mh} (the region
$10\lesssim\omega\lesssim 30$). Another interpolation formula for
the \CC to the pair production process is derived in
\Refs\cite{Lee2004, Lee2004a} up to terms which are of the order
of $1/\omega$ and (1/$\omega^{2}$)$\,\log\omega/2$:
\begin{align}
& \sigma_{C}\left(  \omega\gg2\right) =\alpha\left(  \alpha
Z\right)
^{2}\left\{  -\frac{28}{9}f\left(Z\alpha\right) +\frac{1}{\omega}\left[  -\frac{\pi^{4}}{2}\operatorname{Im}%
g(Z\alpha)\right. \right.  \notag\\
& \left.\left. -4\pi(Z\alpha)^{3}f_{1}(Z\alpha)\right]  +\frac{b}{\omega^{2}}%
\log\frac{\omega}{2}\right\},\label{e5:11}
\end{align}
where the functions $f, g$ and $f_1$ are derived analytically but
the coefficient $b$ is obtained by an interpolation procedure from
the experimental data for $\omega\gtrsim 30$
\Refs\cite{Rosenblum1952, Gimm1978}. The absolute value of the
approximation formula (\ref{e5:11}) is always less than the
corresponding corrections of \Ref\cite{Overbo:1977mh} for
$\omega>25$. The expression (\ref{e5:11}) is zero when
$\omega_0=8.95$ (see \Fig\ref{f5:1}, the corresponding value for
the approximation formula of \Ref\cite{Overbo:1977mh} is
$\omega_0=10.45$). In order to estimate the accuracy of the
integral over the negative part, let us calculate the integral
$-\int_{\omega_0}^\infty d\omega
\sigma_{C}(\omega)/(\omega^{2}\,2\pi^{2}\alpha(
Z\alpha)^{2}f_{2B})$ so that the terms of higher orders in
$1/\omega$ are accounted for in $\sigma_{C}(\omega)$ one after
another. The results are presented in the table (\ref{t5:1}). One
can see that the successive terms from \Eq(\ref{e5:11}) thus
integrated give the contributions of the same order, i.\,e. the
process does not converge to a certain value of the integral.

\begin{table}
\caption{Integration of the $1/\omega$-corrections over the
negative ''tail'' [see \Fig\ref{f5:1} and \Eq({\ref{e5:11}})]
\label{t5:1}}
\begin{ruledtabular}
\begin{tabular}[c]{cc}%
Contribution in $\sigma_C$ when $\omega\to\infty$ &
$\ds-\int_{\omega_0}^\infty d\omega
\frac{\sigma_{C}(\omega)}{\omega^{2}\,2\pi^{2}\alpha(
Z\alpha)^{2}f_{2B}}$ \\ \hline%
$\Oc(1)$ & $-0.184$\\%
$\Oc(1)+\Oc(1/\omega)$ & $0.068$\\%
$\Oc(1)+\Oc(1/\omega)+\Oc(1/\omega^{2})$ & $-0.062$
\end{tabular}
\end{ruledtabular}
\end{table}

It is also quite possible that, in order to resolve the
contradiction between the results (\ref{e5:8}) and (\ref{e5:9}),
the pair production in bound-free states should be taken into
account because of the strong cancellation \Eq(\ref{e5:12}) of the
contribution of free-free states.

\begin{figure}
\includegraphics[width=0.47\textwidth]{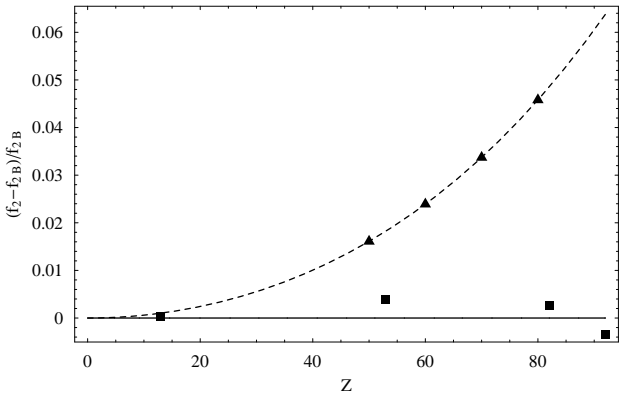}%
\caption{ Our results (triangles) for
$\left(f_{2}(Z)-f_{2B}\right)/f_{2B}$ and the approximation
formula \Eq(\ref{e4:5}) (dashed line) in comparison with the
results of \Ref\cite{Solberg1995} (squares)}.
\label{f5:2}%
\end{figure}

The expression (\ref{e5:9}) coincides with that calculated in
\Ref\cite{Solberg1995} (more precisely, $-D_{1}/f_{2B}$ in the
notations of \Ref\cite{Solberg1995}). The comparison of our
results, i.e. $\left(  f_{2}(Z)-f_{2B}\right)  /f_{2B}$, and the
results of \Ref\cite{Solberg1995} $-D_{1}/f_{2B}$ is made in
\Fig\ref{f5:2}.

It should be noted that our results and those of
\Ref\cite{Solberg1995} are essentially different because the last
one have a non-monotonic dependance on $Z$.

\section{$g$-factor of a bound electron \label{s6}}

\begin{figure}
\begin{center}
\includegraphics[width=0.47\textwidth ]{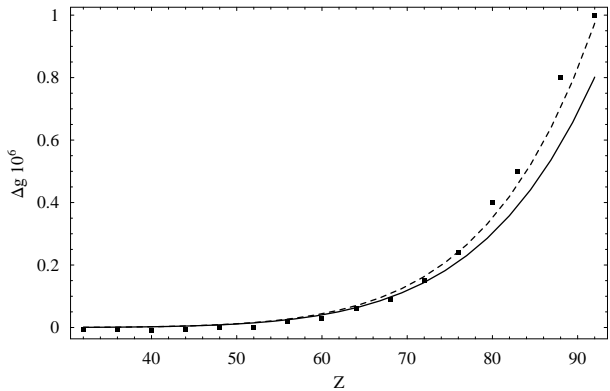}%
\caption{The squares represent the difference $\Delta
g_{\text{Ref.\cite{Beier2000}}}-\Delta
g_{\text{Ref.\cite{Lee:2004vb}}}$, the solid line corresponds to
the function $(16/3)\,\alpha\left( Z\alpha\right)
^{5}(3.35\cdot10^{-4}(Z\alpha)^{2})$, the dashed line is the
function $(16/3)\,\alpha\left(
Z\alpha\right)  ^{5}(3.35\cdot10^{-4}(Z\alpha)^{2}+1.6\cdot10^{-4}%
(Z\alpha)^{4})$, corresponding to the \CC to the form factor
$f_{1}$ \Eq(\ref{e4:4}).\label{f6:1}}
\end{center}
\end{figure}

The amplitude of virtual light-by-light scattering is known to be
a part of the so-called magnetic loop contribution to the
$g$-factor of a bound electron \Ref\cite{Karshenboim:2002jc}.

For the $1S_{1/2}$ electron state, this contribution reads (see
 \Ref\cite{Lee:2004vb}):
\begin{align}
\Delta g & =-\frac{32}{3}\frac{\alpha(Z\alpha)^{2}}{\pi
m^{2}}\int_{0}^{\infty }dq\,f_{1}\left(  q/m\right)\notag\\
& \times \int_{0}^{\infty}dr\,r \tilde{f}_{1}\left(  r\right)
\,\tilde{f}_{2}\left(  r\right) \left( \frac{\sin qr}{qr}-\cos
qr\right),\label{e6:1}
\end{align}
where $\tilde{f}_{1}$ and $\tilde{f}_{2}$ are the radial parts of
the electron wave function in a Coulomb field:
\begin{align}
\psi\left(  \mathbf{r}\right)  =\left(
\begin{array}
[c]{c}%
\tilde{f}_{1}\Omega\\
-i\tilde{f}_{2}\left(  \mathbf{\sigma}\cdot\mathbf{n}\right)
\Omega
\end{array}
\right)  ,\label{e6:2}
\end{align}
where $\Omega$ is the spherical spinor (see for example
\Ref\cite{Berestetskii1982}). Using the \BA\ for the form factor
$f_1$ and the nonrelativistic expressions for the components of
the wave function $\tilde{f}_{1}\left( r\right)=2\exp(-r/a_B)$ and
$\tilde{f}_{2}=\tilde{f}_{1}^{\prime}(r)/2m$ yields the leading
correction to the $g$-factor of a bound electron in $1S_{1/2}$
state \Ref\cite{Karshenboim:2002jc}:
\begin{align}
\Delta g =\frac{7}{216}\text{\,}\alpha(Z\alpha)^{5}.\label{e6:3}
\end{align}
In the case of small $Z\alpha$, one can expand the form factor
$f_{1}$ in power series of $Z\alpha$:
\begin{align}
f_{1}\left(0,0,\mathbf{q},Z\alpha\right)  =F\left(  \frac{q}%
{m}\right)  +\left(  Z\alpha\right)  ^{2}F_{(1)}\left(
\frac{q}{m}\right) +\mathcal{O}\left( Z^{4}\alpha^{4}\right).
\label{e6:1a}
\end{align}
The contribution of $F\left( q/m\right)$ was considered in
\Ref\cite{Lee:2004vb} in detail. To calculate the correction in
$Z\alpha$, it is sufficient to use the functions $\tilde{f}_{1}$
and $\tilde{f}_{2}$ in the nonrelativistic limit and the
expression (\ref{e4:4}) $f_{1}-7/1152$ as $F_{(1)}(0)$. The
results of the numerical calculation of the magnetic-loop
contribution exact in $Z\alpha$ are presented in
\Ref\cite{Beier2000}.

The comparison of the contribution of the \CC to the form factor
$f_1$ \Eq(\ref{e4:4}) and the difference of the results obtained
in \Ref\cite{Beier2000} and \Ref\cite{Lee:2004vb} is depicted in
\Fig\ref{f6:1}. It is surprising that our correction coincides
with this difference not only for the small $Z\alpha$, but for
$Z\alpha\sim1$ also.

\section{Conclusions \label{s7}}

The \CC to  the \Du scattering amplitude have been considered in
this article. We have calculated these corrections in the
low-energy limit but taking into account all orders of the
parameter $Z\alpha$. The accuracy of the calculation is of the
order of one percent for $Z=50$ and increases with $Z$. Our result
is in a good agreement with the corresponding contribution to the
$g$-factor of a bound electron calculated previously in
\Ref\cite{Beier2000}. However, there is a contradiction with the
dispersion integral of the \CC to the pair production cross
section calculated in \Ref\cite{Solberg1995}.

\begin{acknowledgments}
We would like to thank A.I.~Milstein and R.N.~Lee for their
helpful comments and discussions.
\end{acknowledgments}


\appendix*
\section{Example of rearrangement of a conditionally convergent integral \label{a1}}

While calculating the contribution of the first order in
$Z\alpha$, we have expanded the expression (\ref{e2:9}) on
$Z\alpha$ and integrated over $s$. The equation (\ref{e3:12})
corresponding to the contribution of the diagram
[\Fig\ref{f3:1}(b)] has been derived by integration over $\rho$,
$\sigma$ and $t$ in order [see \Eq(\ref{e3:11})]. However, one can
calculate this contribution in an alternative way -- by
substituting the expansion of \Eq(\ref{e2:9}) in \Eq(\ref{e3:4})
and integrating over $r_1$ and $r_2$ before the integration over
$s_1$ and $s_2$ in the Green functions (\ref{e2:9}). One of the
typical expressions appeared is
\begin{multline}
Y\left(  t_{1},t_{2},z\right) =\frac{1+t_{1}t_{2}}{\left(  1+2\,z(t_{1}%
+t_{2})^{2}\right)  ^{2}(t_{1}+t_{2})}\\\times\left(
1-\frac{6(1+t_{1}t_{2})}{\left( t_{1}+t_{2}\right)
^{2}}\right),\label{e7:1}
\end{multline}
where $t_{1,2}=\coth s_{1,2}\in(1,\infty)$ and
$z=(1+x)/2\in(0,1)$. The expression (\ref{e7:1}) has to be
integrated over the total variables' domain. One can easily
integrate over $t_1$, $z$ and $t_2$ one after another (or $z$,
$t_1$ and $t_2$) and obtain a finite result, that is
\begin{align}
&  \int_{1}^{\infty}\int_{0}^{1}\int_{1}^{\infty}Y\,dt_{1}\,dz\,dt_{2}%
=\int_{1}^{\infty}\int_{1}^{\infty}\int_{0}^{1}Y\,dz\,dt_{1}dt_{2}\notag\\
&  =\frac{133}{60}-\frac{13\pi}{4\sqrt{2}}+\frac{119\arctan2\sqrt{2}}%
{60\sqrt{2}}+\frac{38}{15}\,\log \frac{32}{81}.
\end{align}
However, if one integrates \Eq(\ref{e7:1}) over $t_1$ and $t_2$ at
first then the result is the function of $z$
\begin{align}
& \tilde{Y}\left(  z\right) =\int_{1}^{\infty}\left(
\int_{1}^{\infty
}Y\left(  t_{1},t_{2},z\right)  \,dt_{1}\,\right)  dt_{2}\notag\\
& =\frac{1}{60 z}\left\{\frac{1+120z}{2\sqrt{2z}}\left(
\arctan\left(
2\sqrt{2z}\right)  -\frac{\pi}{2}\right)  \right.  \notag\\
& \left. + 16z( 5-48z)  \log\left( \frac{1+8z}%
{8z}\right)  +96z-1  \right\}  ,
\end{align}
which has a nonintegrable singularity at $z=0$
\begin{align}
\tilde{Y}\left(  z\right)
=-\frac{\pi}{240\sqrt{2}z^{3/2}}+O\left( z^{-1/2}\right).
\end{align}

\bibliographystyle{apsrev}
\bibliography{texteng}

\end{document}